\documentclass[twocolumn,showpacs,preprintnumbers,amsmath,amssymb]{revtex4}

\usepackage{graphicx}

\usepackage{amsmath}

\usepackage{amsfonts}

\usepackage{amssymb}

\usepackage{bm}

\newcommand{\ve}{\mathbf}

\begin{document}

\title{Multi Shell Model for Majumdar-Papapetrou Spacetimes}

\author{Metin G\"{u}rses}

 \affiliation{Mathematics Department, Bilkent University, 06800 Ankara-Turkey}

\author{Burak Himmeto\={g}lu}

 \affiliation{Physics Department, Bilkent University, 06800 Ankara-Turkey}

\begin{abstract}

Exact solutions to static and non-static Einstein-Maxwell
equations in the presence of extremely charged dust embedded on
thin shells are constructed. Singularities of multi-black hole
Majumdar- Papapetrou and Kastor-Traschen solutions are removed by
placing the matter on thin shells. Double spherical thin shell
solution is given as an illustration and the matter densities on
the shells are derived.
\end{abstract}


\email{gurses@fen.bilkent.edu.tr}

\email{himmet@ug.bilkent.edu.tr}

\maketitle

\section{Introduction}

It was shown  that the Majumdar-Papapetrou (MP)
solution~\cite{Maj}-~\cite{Pap} describes a spacetime possessing
$N$ extremely charged black holes~\cite{HH}. Later, extremely
charged static dust sources for MP spacetimes were
considered~\cite{MG1}-~\cite{MG2} and it was shown that solutions
to Einstein-Maxwell equations with extremely charged dust
restricted to thin shells exist. Moreover, solutions for various
geometries including planar, spherical and cylindrical shells were
obtained~\cite{MG2}. It was also shown that spherical shells can
be used as sources for the Extreme Reissner-N\"{o}rdstrom (ERN)
spacetimes~\cite{MG2}. The cosmological black-hole solutions which
are time dependent generalizations of the MP solution were also
studied~\cite{KT}-~\cite{BHKT} and generalizations with
discussions on non-radiative character of these spacetimes were
discussed in~\cite{B}.

In linear theories mass density of point particle distribution can
be expressed as
$\rho(\ve{r})=\sum_{i}m_{i}\delta(\ve{r}-\ve{a_{i}})$ where
$m_{i}$'s are the masses of each particle, $\delta(\ve{r})$ is the
Dirac delta function and $a_{i}$'s are the locations of the point
particles. This expression is in consistence with the Poisson's
equation,
\begin{equation}\label{pois}
\nabla^{2}V=-4\pi\rho \quad \mathrm{with} \quad
V=\sum_{i=1}^{N}\frac{m_{i}}{|\ve{r}-\ve{a_{i}}|}.
\end{equation}
Such a point particle representation of the mass density is not
valid in nonlinear theories. For instance, in a nonlinear theory
like $\nabla^{2}V+4{\pi\rho}V^3=0$, there exists no solution $V$
for $\rho=\sum_{i}m_{i}\delta(\ve{r}-\ve{a_{i}})$. We overcome the
above problem by replacing massive particles by spherical thin
shells with centers at $\ve{r}=\ve{a_{i}}$ and radii $r_{i}$. Then
the mass density $\rho$ can now be represented as
$\rho(\ve{r})=\sum_{i=1}^{N}\rho_{0i}\delta(F_{i})$ where
$F_{i}(\ve{r})=0$ $(i=1,2,..,N)$ represents the positions of the
shells and $\rho_{0i}$ are the mass density of each thin shell.
The space is divided into $N+1$ spaces. In our proposed method we
shall solve such potential problems where the solutions are
continuous everywhere.

In this work, we will first review the solutions to
Einstein-Maxwell equations in the presence of extremely charged
static dusts restricted to thin shells that are given
in~\cite{MG2}. Then we will generalize the methods in~\cite{MG2}
to find solutions possessing extremely charged dusts restricted to
multiple shells and obtain singularity free versions of
MP-solutions given in~\cite{HH}. We will also work out the double
shell problem explicitly and show that the matter distributions on
the shells are not uniform and that interior of the shells are not
flat unlike the single shells in~\cite{MG2}. Finally we will work
on cosmological MP solutions discussed
in~\cite{KT}-~\cite{BHKT}-~\cite{B} and develop similar multi
shell models.

\section{Shell Models for the Majumdar-Papapetrau Spacetimes}
Let $\mathcal{M}$ be a four dimensional spacetime with the metric,
\begin{equation}\label{MPmetric}
ds^2=-\lambda^{-2}dt^2+\lambda^{2}h_{ij}dx^{i}dx^{j},
\end{equation}
where $h_{ij}$ is an Euclidean 3-metric and $\lambda$  is a
function of spatial coordinates $x^{i}$ only. With this metric
assumption, we consider a charged static dust source with the
matter part of the energy-momentum tensor given by,
\begin{equation}\label{T}
T_{\mu\nu}^{M}={\rho}u_{\mu}u_{\nu},
\end{equation}
where $u_{\mu}=-\frac{1}{\lambda}\delta_{\mu}^{0}$ since the dust
is static. The four-potential due to the static charged dust is in
the form $A_{\mu}=A_{0}(x^{i})\delta_{\mu}^{0}$, so the current
four-vector is given by,
\begin{equation}\label{J}
J^{\mu}=\sigma(x^{i})u^{\mu}=\sigma(x^{i})\lambda\delta_{\mu}^{0},
\end{equation}
where $\sigma(x^{i})$ is the charge density of the dust. Therefore
the electromagnetic and the Maxwell energy-momentum tensors are
given by,
\begin{eqnarray}
F_{\mu\nu} &=& \nabla_{\nu}A_{\mu}-\nabla_{\mu}A_{\nu} \label{F},
\\
M_{\mu\nu} &=& \frac{1}{4\pi}(F_{\mu\alpha}F_{\nu}^{\alpha}-
\frac{1}{4}F_{\alpha\beta}F^{\alpha\beta}g_{\mu\nu}). \label{M}
\end{eqnarray}
Using the Einstein tensor calculated from the metric
(\ref{MPmetric}) and the field equations
$G_{\mu\nu}=8\pi(T_{\mu\nu}^{M}+M_{\mu\nu})$ with
(\ref{T})-(\ref{M}) we get,
\begin{equation}\label{A}
A_{0}=\frac{\kappa}{\lambda}, \quad \kappa={\pm}1 .
\end{equation}
Then plugging the above equation back into the Einstein equations
give,
\begin{equation}\label{Nonlin}
\nabla^{2}\lambda+4\pi\rho\lambda^{3}=0.
\end{equation}
Therefore the Einstein-Maxwell equations reduce to a nonlinear
type of Poisson's equation. To find the equation satisfied by the
charge density $\sigma(x^{i})$ we use the Maxwell equations
$\nabla_{\nu}F^{\mu\nu}=4{\pi}J^{\mu}$ with (\ref{J}) and
(\ref{F}) to get,
\begin{equation}\label{Nonlin-2}
\nabla^{2}\lambda+4\pi\kappa\sigma\lambda^{3}=0.
\end{equation}
Comparing (\ref{Nonlin}) and (\ref{Nonlin-2}) gives
$\rho=\kappa\sigma$ which shows that the dust source is extremely
charged.

One can solve the reduced Einstein-Maxwell equations
(\ref{Nonlin})-(\ref{Nonlin-2}) for the extremely charged dust
restricted to a thin shell as given in~\cite{MG2}. Letting $S$ be
a regular surface in $\mathbb{R}^{3}$ defined by
$S=[(x,y,z)\in\mathbb{R}^{3};F(x,y,z)=0]$, the matter density can
be written as,
\begin{equation}\label{rho-single}
\rho(\ve{r})=\rho_{0}(\ve{r})\delta(F(\ve{r})),
\end{equation}
where $\rho_{0}$ is the matter distribution on the shell $S$.
Generalizing the method given in~\cite{MG2} we can set,
\begin{equation}\label{lambda-single}
\lambda(\ve{r})=\lambda_{0}(\ve{r})-\lambda_{1}(\ve{r})\Theta(F)-
\lambda_{2}(\ve{r})(1-\Theta(F)),
\end{equation}
where $\Theta(F)$ is the Heaviside step function. The above choice
can be made since the spacetime $\mathcal{M}$ is now divided into
two disjoint spacetimes  $\mathcal{M}^{+}$ and $\mathcal{M}^{-}$
by $S$ where the metric functions $\lambda$ are
$\lambda_{0}-\lambda_{1}$ and $\lambda_{0}-\lambda_{2}$
respectively. We also require that the following equations are
satisfied,
\begin{eqnarray}
\nabla^{2}\lambda_{0} = \nabla^{2}\lambda_{1} &=&
\nabla^{2}\lambda_{2} = 0, \label{Cond-1} \\
(\lambda_{1}-\lambda_{2})|_{S} &=& 0. \label{Cond-2}
\end{eqnarray}
These follow from the fact that in $\mathcal{M}^{+}$ and
$\mathcal{M}^{-}$ the Laplace's equation is satisfied as can be
seen from (\ref{Nonlin})(Since $\rho=0$ in $\mathcal{M}^{+}$ and
$\mathcal{M}^{-}$) and the metric must be continuous across the
shell. Inserting (\ref{lambda-single}) into (\ref{Nonlin}) and
using the conditions (\ref{Cond-1})-(\ref{Cond-2}) we get,
\begin{equation}\label{rho0-single}
\rho_{0}=\frac{({\nabla}\lambda_{1}-
{\nabla}\lambda_{2}){\cdot}{\nabla}F}{4{\pi}\lambda_{0}^{3}}|_{S}.
\end{equation}
Let $S$ be the sphere $F=r-a=0$. One can choose $\lambda_{2}=0$,
$\lambda_{1}=\lambda_{3}(\theta,\phi)(\frac{1}{a}-\frac{1}{r})$
which clearly satisfies (\ref{Cond-1})-(\ref{Cond-2}). Using
(\ref{rho0-single}) gives,
\begin{equation}\label{rho0-sphere}
\rho_{0}(\theta,\phi)=\frac{\lambda_{3}(\theta,\phi)}{4{\pi}a^2(\lambda_{0}|_{S})^{3}},
\end{equation}
where $\lambda_{3}(\theta,\phi)$ satisfies,
\begin{equation}\nonumber
\left[\frac{1}{r^2\sin{\theta}}\frac{\partial}{\partial\theta}\left(\sin{\theta}
\frac{\partial}{\partial\theta}\right)+\frac{1}{r^2\sin^2{\theta}}
\frac{\partial^2}{\partial\phi^2}\right]\lambda_{3}(\theta,\phi)=0.
\end{equation}
At this point we would like to remark that the above equation
(\ref{rho0-sphere}) can also be obtained by using the Israel
junction conditions~\cite{I}. However, when Israel junction
conditions are used the surface energy-momentum tensor has the
components,
\begin{equation}\label{S}
S_{tt}=\frac{\lambda_{3}(\theta,\phi)}{4{\pi}a^2(\lambda_{0}|_{S})^{4}},
\quad S_{\theta\theta}=S_{\phi\phi}=0.
\end{equation}
So the matter density $\rho_{I}$ on $S$ is given by using the fact
that
$S_{tt}={\rho_{I}}u_{t}u_{t}=\frac{\rho_{I}}{\lambda_{0}^{2}}|_{S}$,
\begin{equation}\label{rh0-Israel}
\rho_{I}(\theta,\phi)=\frac{\lambda_{3}(\theta,\phi)}{4{\pi}a^2(\lambda_{0}|_{S})^{2}}.
\end{equation}
As can be seen from (\ref{rh0-Israel}) in Israel method we get
$\lambda_{0}^{2}$ in the denominator instead of $\lambda_{0}^{3}$
as in (\ref{rho0-sphere}) which is due to a dimensional scaling.
The total mass on the shell can be written as an integral of
$\rho_{0}\delta(F)$ over the 3 spatial dimensions which is equal
to the integral of $\rho_{I}$ over the 2 dimensional surface, that
is,
\begin{equation}\label{remark}
\int_{\mathcal{V}_{3}}\rho_{0}\delta(F)\,
\lambda^3\,d^3x=\int_{\mathcal{S}_{2}}\rho_{I}\lambda^2d^2x,
\end{equation}
So we get the same total mass on the shell in either method which
means that there is no ambiguity.

When the interior of the sphere $S$ is chosen as a flat spacetime,
we can set $\lambda_{0}=1$ and $\lambda_{3}=m_{0}$ as constant
which turns out to be the mass of the shell. So
(\ref{rho0-sphere}) becomes,
\begin{equation}\label{rho-ERN-single}
\rho_{0}=\frac{m_{0}}{4{\pi}a^2}.
\end{equation}
So the interior and exterior metric functions are given by,
\begin{eqnarray}
\lambda_{out} &=& 1-\frac{m_{0}}{a}+\frac{m_{0}}{r},
\label{lambda-out-single} \\
\lambda_{in} &=& 1, \label{lambda-in-single}
\end{eqnarray}
where $m_{0}$ is the mass of the shell as can clearly be seen from
(\ref{rho-ERN-single}). This solution represents the ERN solution
exterior to a spherical shell with flat interior. Letting
$r=(R-m_{0})/{\beta}$ where $\beta{\equiv}1-m_{0}/a$ with
$m_{0}{\neq}a$ assumed, the exterior metric can be written as,
\begin{equation}\label{metric-ext-single}
ds^2=-\beta^2(1-\frac{m_{0}}{R})^2dt^2+(1-\frac{m_{0}}{R})^2dR^2+R^2d\Omega^2,
\end{equation}
which is the conventional form of ERN metric (after a scaling). As
we noted earlier, we aim to remove the singularity of an ERN
spacetime. Choosing $m_{0}<a$ we see from
(\ref{lambda-out-single}) that $\lambda_{out}>0$ is always
satisfied. Comparing (\ref{metric-ext-single}), the relation
$r=(R-m_{0})/{\beta}$ and (\ref{lambda-out-single}) one can
realize that the singularity of ERN spacetime is at
$\lambda_{out}=0$ (which corresponds to $R=0$) and the horizon is
at $r=0$ (which corresponds to $R=m_{0}$). This means that by
restricting the matter on the shell with $m_{0}<a$ we remove the
both the singularity and the horizon of the ERN spacetime. In the
case $m_{0}>a$ (\ref{lambda-out-single}), $\lambda_{out}=0$ is not
excluded. The case $m_{0}=a$ represents the
Levi-Civita-Bertotti-Robinson (LCBR) spacetime outside the shell
and flat spacetime inside. By letting $r=m_{0}^2/R$ the exterior
metric is obtained as,
\begin{equation}\label{LCBR}
ds^2=\frac{m_{0}}{R^2}[-dt^2+dR^2+R^2d\Omega^2],
\end{equation}
as the usual conformally flat LCBR metric. At this point we would
like to note that in our analysis the limiting case
$a{\rightarrow}0$ does not exist since in that limit the matter
density (\ref{rho-single}) would become
$\rho(\ve{r})=\rho_{0}(\ve{r})\delta(\ve{r})$ and this choice is
inconsistent with the Einstein-Maxwell equations
(\ref{Nonlin})-(\ref{Nonlin-2}). If we had the Newtonian theory,
such a limit mathematically  would be consistent with the
Poisson's equation. The non-existence of the $a{\rightarrow}0$
limit is in agreement with the results of~\cite{GT}.

To remove the singularities of a MP spacetime possessing $N$ ERN
black holes discussed in~\cite{HH}, we place  the matter source on
thin shells. A MP spacetime possessing $N$ ERN black holes has the
metric function $\lambda$ given by,
\begin{equation}\label{lambda-N-BH}
\lambda=1+\sum_{j=1}^{N}\frac{m_{i}}{|\ve{r}-\ve{a_{i}}|},
\end{equation}
where $\ve{a_{i}}$ is the position of the $i^{th}$ ERN black hole.
Such a spacetime contains $N$ number of singularities. We can
place the extremely charged dust on $N$ spatially separated shells
that do not intersect, instead of considering point sources.
Generalizing the choice we made in (\ref{rho-single}) we can write
the matter density of the spacetime as,
\begin{equation}\label{rho-multi}
\rho(\ve{r})=\sum_{j=1}^{N}\rho_{0j}(\ve{r})\delta(F_{j}),
\end{equation}
where $\rho_{0j}$ is the matter distribution on the $j^{th}$ shell
defined by $S_{j}:F_{j}(\ve{r})=0$. We can also generalize the
choice for $\lambda$ as,
\begin{equation}\label{lambda-multi}
\lambda=\lambda_{0}-\lambda_{e}\prod_{j=1}^{N}\Theta(F_{j})-
\sum_{j=1}^{N}\lambda_{j}[1-\Theta(F_{j})],
\end{equation}
so that the metric function inside the $j^{th}$ shell is given by,
\begin{equation}\label{lambda-j}
\lambda_{j}^{in}=\lambda_{0}-\lambda_{j} \quad {\forall}j=1,..,N.
\end{equation}
Moreover the metric function exterior to all of the $N$ shells is
given by,
\begin{equation}\label{lambda-ext}
\lambda^{ext}=\lambda_{0}-\lambda_{e}.
\end{equation}
The functions $\lambda_{0}$, $\lambda_{j}$ and $\lambda_{e}$
satisfy the following,
\begin{eqnarray}
\nabla^{2}\lambda_{0}=\nabla^{2}\lambda_{j}=
\nabla^{2}\lambda_{e} &=& 0, \label{Cond-multi-1} \\
(\lambda_{e}-\lambda_{j})|_{S_{j}}=0 \label{Cond-multi-2} \quad
{\forall}j=1,..,N,
\end{eqnarray}
where we again use the fact that the Laplace's equation is
satisfied in source free regions and the metric across the shells
must be continuous. Using
(\ref{lambda-multi})-(\ref{Cond-multi-1}) and (\ref{Cond-multi-2})
we calculate $\nabla^{2}\lambda$ as,
\begin{equation}\label{Laplacian-multi}
\nabla^{2}\lambda=\sum_{j=1}^{N}(\nabla\lambda_{j}-
\nabla\lambda_{e})\cdot\nabla{F_{j}}\delta(F_{j}).
\end{equation}
Using the fact that $\delta(F_{j})\Theta(F_{k})=\delta(F_{j})$ for
$k{\neq}j$, $\rho\lambda^3$ is given by,
\begin{equation}\label{rho-lambda-3}
\rho\lambda^3=\sum_{j=1}^{N}\rho_{0j}(\lambda_{0}-\lambda_{j})^3\delta(F_{j}).
\end{equation}
Then inserting (\ref{Laplacian-multi}) and (\ref{rho-lambda-3})
into (\ref{Nonlin}) we get,
\begin{equation}\label{rho-0-multi}
\rho_{0j}=\frac{(\nabla\lambda_{e}-
\nabla\lambda_{j})\cdot\nabla{F_{j}}}{4\pi(\lambda_{0}-\lambda_{j})^3}|_{S_{j}}.
\end{equation}
As an illustration we consider the case of two spherical shells
with radii $r_{1}$ and $r_{2}$ and the centers located at
$\ve{a_{1}}$ and $\ve{a_{2}}$. Our aim is to obtain the exterior
MP solution with the metric function given in (\ref{lambda-N-BH})
so that the singularities can be removed. The equations defining
the spherical shells are given by
$F_{1}=|\ve{r}-\ve{a_{1}}|-r_{1}=0$ and
$F_{2}=|\ve{r}-\ve{a_{2}}|-r_{2}=0$. Making analogy with
(\ref{lambda-N-BH}) and the single shell solution
(\ref{lambda-out-single}) we choose,
\begin{equation}\label{lambda-e-multi}
\lambda_{e}=\frac{m_{1}}{r_{1}}+\frac{m_{2}}{r_{2}}-\frac{m_{1}}{|\ve{r}-\ve{a_{1}}|}-
\frac{m_{2}}{|\ve{r}-\ve{a_{2}}|}.
\end{equation}
Appropriate choices for $\lambda_{1}$ and $\lambda_{2}$ can be
made by the following forms using (\ref{Cond-multi-1}) and
(\ref{Cond-multi-2}),
\begin{eqnarray}
\lambda_{1} &=&
\frac{m_{2}}{r_{2}}-\frac{m_{2}}{|\ve{r}-\ve{a_{2}}|},
\label{lambda-1-multi} \\
\lambda_{2} &=&
\frac{m_{1}}{r_{1}}-\frac{m_{1}}{|\ve{r}-\ve{a_{1}}|},
\label{lambda-2-multi}
\end{eqnarray}
which in turn gives the full metric functions with the choice
$\lambda_{0}=1$ as,
\begin{eqnarray}
\lambda^{ext} &=&
1-\frac{m_{1}}{r_{1}}-\frac{m_{2}}{r_{2}}+\frac{m_{1}}{|\ve{r}-
\ve{a_{1}}|}+\frac{m_{2}}{|\ve{r}-\ve{a_{2}}|},
\label{lambda-ext-multi} \\
\lambda_{1}^{in} &=&
1-\frac{m_{2}}{r_{2}}+\frac{m_{2}}{|\ve{r}-\ve{a_{2}}|},
\label{lambda-in-1} \\
\lambda_{2}^{in} &=&
1-\frac{m_{1}}{r_{1}}+\frac{m_{1}}{|\ve{r}-\ve{a_{1}}|}.
\label{lambda-in-2}
\end{eqnarray}
The above choices describe an exterior MP spacetime and ERN
spacetimes inside each shell. Let $\ve{a_{1}}=-a\ve{e_{z}}$ and
$\ve{a_{2}}=a\ve{e_{z}}$ so that the centers of the shells are on
the z-axis with $r_{1}+r_{2}<2a$. The surface matter distributions
are obtained from (\ref{rho-0-multi}) with $N=2$ as,
\begin{eqnarray}
\rho_{01} &=&
\frac{m_{1}}{4{\pi}r_{1}^2}\left[1-\frac{m_{2}}{r_{2}}+
\frac{m_{2}}{\sqrt{r_{1}^2-4ar_{1}\cos{\theta_{1}}+4a^2}}\right]^{-3},
\label{rho-2-sphere-1} \\
\rho_{02} &=&
\frac{m_{2}}{4{\pi}r_{2}^2}\left[1-\frac{m_{1}}{r_{1}}+
\frac{m_{1}}{\sqrt{r_{2}^2+4ar_{2}\cos{\theta_{2}}+4a^2}}\right]^{-3},
\label{rho-2-sphere-2}
\end{eqnarray}
where $\theta_{1}$ and $\theta_{2}$ are spherical coordinates on
$S_{1}$ and $S_{2}$ with their centers are taken as if origin. It
can clearly be seen from (\ref{rho-2-sphere-1}) and
(\ref{rho-2-sphere-2}) that the existence of a second shell
disturbs the uniform matter distribution on the other shell so
that the system stays at equilibrium. The above equations can be
obtained by Israel method as we discussed before, where one
obtains the second power of the term in the denominator instead of
the third power as in (\ref{rho-2-sphere-1}) and
(\ref{rho-2-sphere-2}) which is due to dimensional scaling. One
can obtain (\ref{lambda-N-BH}) by letting
$\beta{\equiv}1-\frac{m_{1}}{r_{1}}-\frac{m_{2}}{r_{2}}$ in
(\ref{lambda-ext-multi}) so $\lambda^{ext}$ becomes,
\begin{equation}\label{lambda-ext-multi*}
\lambda^{ext}=\beta\left(1+\frac{\bar{m}_{1}}{|\ve{r}-
\ve{a_{1}}|}+\frac{\bar{m}_{2}}{|\ve{r}-\ve{a_{2}}|}\right)=\beta\bar{\lambda}_{ext},
\end{equation}
where $\bar{m}_{1}=m_{1}/{\beta}$ and $\bar{m}_{2}=m_{2}/{\beta}$.
The MP spacetime possessing two ERN black holes has two horizons
each described as a surface enclosing the points where the black
holes are located~\cite{HH}. These singularities are defined by
the vanishing of $\lambda^{ext}$. So choosing the masses and the
radii of the shells such that $m_{1}/r_{1}+m_{2}/r_{2}<1$, we
guarantee that $\lambda^{ext}>0$ from (\ref{lambda-ext-multi}),
which means that by this choice we remove the singularities. Note
that $\lambda_{1}^{in}$ in (\ref{lambda-in-1}) and
$\lambda_{2}^{in}$ in (\ref{lambda-in-2}) can never vanish in
their domains of definition. The case $m_{1}/r_{1}+m_{2}/r_{2}>1$
is irrelevant since we can not remove the singularities of the MP
spacetime. The case $m_{1}/r_{1}+m_{2}/r_{2}=1$ should correspond
to two mass generalization of the LCBR metric. Thus we obtained
the double ERN black hole solutions with singularities removed.
(Note that we obtained the solution with a re-scaling factor
$\beta$). At this point we remark that MP-solutions with multiple
spherical dust shells as sources, can not have flat interiors as
we showed in (\ref{lambda-in-1}) and (\ref{lambda-in-2}).
Moreover, the shells disturb each other and cause angular
dependence of matter density on the shells. One can see from
(\ref{rho-2-sphere-1}) and (\ref{rho-2-sphere-2}) that the second
shell($S_{2}$) has maximum matter density at its north pole and
the first shell($S_{1}$) has its maximum matter density at its
south pole. Such a configuration puts the system in equilibrium.
One can also calculate the masses of the shells by integrating
(\ref{rho-2-sphere-1}) and (\ref{rho-2-sphere-2}) on $S_{1}$ and
$S_{2}$ respectively as in (\ref{remark}) which gives the mass of
the first shell as $m_{1}$ and the second shell as $m_{2}$ which
is expected. We again note that the limits $r_{1}{\rightarrow}0$
and $r_{2}{\rightarrow}0$ do not exist since the matter density
(\ref{rho-multi}) is not consistent with the Einstein-Maxwell
equations (\ref{Nonlin})-(\ref{Nonlin-2}). This means that we can
not have a source density of the form,
\begin{equation}\nonumber
\rho(\ve{r})=m_{1}\delta(\ve{r}-\ve{a_{1}})+m_{2}\delta(\ve{r}-\ve{a_{2}}).
\end{equation}
This is again in total agreement with~\cite{GT}.

One can also consider other solutions to the Laplace's equation
for the metric function $\lambda$ to obtain various spacetimes. It
was shown that such spacetimes possess naked
singularities~\cite{HH}. One can remove these naked singularities
by placing the extremely charged source on thin shells as we did
in this work for multiple ERN black hole solutions. The same
procedure can also be applied to the stationary generalization of
MP-spacetimes given by Israel-Wilson~\cite{IW} and
Perj\'{e}s~\cite{P} to remove the naked singularities of these
spacetimes.

\section{Shell Models for the Kastor-Traschen Spacetimes}

It was shown that time-dependent generalizations of MP spacetimes
exist and solutions corresponding to $N$ extremely charged
co-moving black holes in a de-Sitter background were considered
in~\cite{KT}-~\cite{BHKT}-~\cite{B}. The idea of removing the
singularities of $N$ co-moving black hole solutions by thin shells
were considered briefly in~\cite{BHKT} for testing the cosmic
censorship conjecture. At this point we would like to extend the
discussion given in~\cite{BHKT} by the methods we discussed above.

The cosmological MP solution which is a time-dependent
generalization of the MP solution with metric (\ref{MPmetric}) in
cosmological coordinates is given by,
\begin{equation}\label{MPcosmo}
ds^2=-\tilde{U}^{-2}dt^2+R(t)^2 \, \tilde{U}^2h_{ij}dx^{i}dx^{j},
\end{equation}
where $R(t)$ is the scale factor and $h_{ij}$ is an Euclidean
metric. For a single extreme Reissner-N\"{o}rdstrom-de-Sitter
(ERNdS) black hole with its charge equal to its mass $(Q=M)$ the
metric function $\tilde{U}$ in the metric (\ref{MPcosmo}) is given
by,
\begin{equation}
\tilde{U}=1+e^{-Ht}\,{M \over r}.
\end{equation}
We can obtain the ERNdS solution in static coordinates by
following the transformation given in~\cite{KT} Given the metric
(\ref{MPcosmo}) with $\tilde{U}$ being arbitrary we consider a
co-moving charged dust as source. As before, we assume the
four-potential of the form $A_{\mu}=A_{0}\delta_{\mu}^{0}$ and
$u_{\mu}=-\frac{1}{\tilde{U}}\delta_{\mu}^{0}$ since the dust is
co-moving (i.e. static in cosmological coordinates). Then the
Einstein equations with positive cosmological constant become,
\begin{equation}\label{Eins-cosmo}
G_{\mu\nu}=8\pi({\rho}u_{\mu}u_{\nu}+M_{\mu\nu})+{\Lambda}g_{\mu\nu},
\end{equation}
where $M_{\mu\nu}$ is the Maxwell tensor given as in (\ref{M}).
The the Einstein equations (\ref{Eins-cosmo}) give,
\begin{eqnarray}
(\frac{\dot{R}}{R})^2 &=& \frac{\Lambda}{3}, \label{Friedm}
\\
\nabla^{2}\tilde{U} &+& 4{\pi\rho}R(t)^2\tilde{U}^{3}=0,
\label{Nonlin-cosmo}
\\
\tilde{U}&=&1+{1 \over R(t) }\, \lambda (x), \label{lam}
\end{eqnarray}
where dot over $R$ represents derivative with respect to $t$ and
$\lambda(x)$ is independent of $t$. The above equations are
obtained by separating the cosmological and electromagnetic-matter
parts of the Einstein equations. One can note that (\ref{Friedm})
is just the Friedmann equation for a flat cosmology. Solving
(\ref{Friedm}) yields $R(t)=e^{Ht}$ where
$H={\pm}\sqrt{3/\Lambda}$. As discussed
in~\cite{KT}-~\cite{BHKT}-~\cite{B} negative $H$ corresponds to
black hole spacetimes while positive $H$ corresponds to white hole
spacetimes. We consider negative $H$ for the rest of the paper
($H=-|H|$). The Maxwell equations
$\nabla_{\nu}F^{\mu\nu}=4{\pi}J^{\mu}$ give,
\begin{equation}\label{Nonlin-cosmo-2}
\nabla^{2}\tilde{U}+\frac{4\pi\sigma}{\kappa}R(t)^2\tilde{U}^{3}=0,
\end{equation}
with $\kappa={\pm}1$ and $\sigma$ being the charge density of the
co-moving dust. Comparing (\ref{Nonlin-cosmo}) and
(\ref{Nonlin-cosmo-2}) we get $\rho=\kappa\sigma$ as before, so we
conclude that the dust is extremely charged. Letting
$d\tau=R(t)dt$ we can write the metric (\ref{MPcosmo}) as,
\begin{equation}\label{MPcosmo-2}
ds^2=-U^{-2}d{\tau}^2+U^2h_{ij}dx^{i}dx^{j}.
\end{equation}
where
\begin{eqnarray}
U&=&H\tau+\lambda(x) \label{U}, \\
\nabla^2U&+&4\pi\rho{U}^3=0. \label{Nonlin-U}
\end{eqnarray}
One can see from the Einstein equations (\ref{Nonlin-U}) that the
product ${\rho}U^3$ must be time independent. This is clear from
the fact $\nabla^{2}U$ is time independent as can be seen from
(\ref{U}) and to have the equation (\ref{Nonlin-U}) satisfied for
all times we must have ${\rho}U^3$ to be time independent.

The above derivation of Einstein-Maxwell equations for a charged
dust was  given in~\cite{BHKT}. Our treatment of the shell model
will be parallel to which we considered in the previous part of
this work, that is different from the shells considered
in~\cite{BHKT}. We start with a dust model and then take its limit
to thin shells with the use of Dirac delta functions for the
matter density. Since (\ref{Nonlin-cosmo}) and
(\ref{Nonlin-cosmo-2}) are totally same with equations
(\ref{Nonlin}) and (\ref{Nonlin-2}) we can consider similar
multi-shell models with the matter density given in
(\ref{rho-multi}) and the metric function $U$ as in
(\ref{lambda-multi}) (Clearly we just replace $\lambda$ with $U$).
One difference is that the metric function $U$ is time dependent
via $H\tau$. However, as can be seen from (\ref{U}) and
(\ref{Nonlin-U}) this time dependence does not affect
Einstein-Maxwell equations. In source free regions we again have
the Laplace's equation satisfied. The equation of each surface
defining the shells are again given by $F_{j}(\ve{r})=0$. Thus we
follow the lines (\ref{rho-multi}) to (\ref{rho-0-multi}) to get
the same dependence for the surface matter density.

As an example, we again consider two spherical shells with radii
$r_{1}$, $r_{2}$ and with centers located at $\ve{a_{1}}$,
$\ve{a_{2}}$.The equations describing the shells are given by,
\begin{eqnarray}
F_{1} &=& |\ve{r}-\ve{a_{1}}|-r_{1}=0, \nonumber \\
F_{2} &=& |\ve{r}-\ve{a_{2}}|-r_{2}=0. \label{shells-cosmo}
\end{eqnarray}
Due to linearity of Laplace's equation, we extend the single ERNdS
solution in cosmological coordinates given in (\ref{MPcosmo-2}) to
the double ERNdS case for the exterior metric function as,
\begin{equation}\label{U-ext}
U^{ext}=H\tau-\frac{m_{1}}{r_{1}}-\frac{m_{2}}{r_{2}}+\frac{m_{1}}{|\ve{r}-
\ve{a_{1}}|}+\frac{m_{2}}{|\ve{r}-\ve{a_{2}}|},
\end{equation}
where we have chosen $U_{0}=H\tau$ in (\ref{lambda-ext}). Then
(\ref{lambda-in-1}) and (\ref{lambda-in-2}) suggests,
\begin{eqnarray}
U_{1}^{in} &=&H\tau-\frac{m_{2}}{r_{2}}+\frac{m_{2}}{|\ve{r}-\ve{a_{2}}|},
\label{U-in-1} \\
U_{2}^{in} &=&
H\tau-\frac{m_{1}}{r_{1}}+\frac{m_{1}}{|\ve{r}-\ve{a_{1}}|},
\label{U-in-2}
\end{eqnarray}
for the metrics inside the first and the second shells. Clearly
the choices (\ref{U-ext})-(\ref{U-in-1})-(\ref{U-in-2}) satisfy
the required conditions (\ref{Cond-multi-1}) and
(\ref{Cond-multi-2}) for $N=2$. We have mentioned above that
${\rho}U^3$ must be time independent which is satisfied in our
multi-shell model since this product is given with the use of
(\ref{rho-multi}), (\ref{lambda-multi}), (\ref{U-in-1}) and
(\ref{U-in-2}) as,
\begin{equation}\nonumber
{\rho}U^3=\rho_{01}(U_{1}^{in})^3\delta(F_{1})+\rho_{02}(U_{2}^{in})^3\delta(F_{2}).
\end{equation}
Since $F_{1,2}$ are time independent and $\rho_{01,2}U_{1,2}^{in}$
will be shown to be time independent below.

Again choosing $\ve{a_{1}}=-a\ve{e_{z}}$ and
$\ve{a_{2}}=a\ve{e_{z}}$ such that $r_{1}+r_{2}<2a$ the surface
matter distributions are calculated from (\ref{rho-0-multi}) as,
\begin{eqnarray}
\rho_{01} &=&
\frac{m_{1}}{4{\pi}r_{1}^2}\left[H\tau-\frac{m_{2}}{r_{2}}+\frac{m_{2}}{\sqrt{r_{1}^2-
4ar_{1}\cos{\theta_{1}}+4a^2}}\right]^{-3},
\nonumber \\
\rho_{02} &=&
\frac{m_{2}}{4{\pi}r_{2}^2}\left[H\tau-\frac{m_{1}}{r_{1}}+\frac{m_{1}}{\sqrt{r_{2}^2+
4ar_{2}\cos{\theta_{2}}+4a^2}}\right]^{-3}, \nonumber
\end{eqnarray}
where the definitions of $\theta_{1}$ and $\theta_{2}$ are given
as before. As can be seen from above, $\rho_{01}$ and $\rho_{02}$
are time dependent but when they are integrated on the surface of
each shell (the term $\rho_{01,2}U_{1,2}^{in}$ will appear in the
integral which is time independent), they give constant masses for
the shells as $m_{1}$ and $m_{2}$ which is in consistence with the
result of Bonnor ~\cite{B} . If we had a single shell, the
interior of the shell would be de-Sitter. But for the case of
double shells (and for $N>2$ also) as can be seen from
(\ref{U-in-1}) and (\ref{U-in-2}) the interior of the shells are
ERNdS. Unless this choice is made, the continuity condition of the
metric across the shells will be violated. This fact was not
realized in~\cite{BHKT} where the interior of the shells were
considered as de-Sitter.

To remove the singularities of the cosmological multi black hole
solutions we again restrict the matter on thin shells. For that
purpose, we assume that the thin shells are formed at $\tau=0$
with $H<0$ so from (\ref{U-ext}) one can see that $U^{ext}$ never
vanishes for later times $\tau{\geq}0$. This means that the
exterior metric is singularity free for all times. Moreover the
interior of the shells described by (\ref{U-in-1}) and
(\ref{U-in-2}) can also never vanish for the choice we made. Thus
the whole spacetime becomes regular. The spacetime distances
$||PQ||_{g}$  between two neighboring (space) points $P$ and $Q$
is given by $||PQ||_{g}=|U|||P-Q||$ where $||P-Q||$ is the
distance in ${\mathbb R}^3$, and $U$ is given in (\ref{U}). Hence
in exterior and interior spacetimes the space distances increase
with constant (Hubble) speed. For this reason the shells inflate
without collision, because the distances among them also increase.

There is also another interesting possibility of constructing
regular spacetimes by considering $H>0$ for $\tau{\leq}0$ and
$H<0$ for $\tau{\geq}0$. With these choices one can see that
$U^{ext}$, $U_{1}^{in}$ and $U_{2}^{in}$ never vanishes and the
spacetime metric is continuous at $\tau=0$. In this model, the
Hubble constant $H$ changes sign in each universe separated by
$\tau=0$ which causes a delta-function type of singularity at this
hypersurface. Spacetimes (interior and exterior spacetime regions)
are contracting for $\tau\leq{0}$ and expanding for $\tau\geq{0}$.

Our method can be applied also to higher dimensional
generalizations of the Majumdar-Papapetrau-Kastor- Traschen
solutions~\cite{MS} to study brane world creations and
collisions~\cite{Gib}.

\section{Conclusion}

 We considered thin mass shell models for the Majumdar-Papapetrou
 and Kastor-Traschen spacetimes. In both cases we found exact solutions
 of the field equations describing gravitational fields of the extremely charged
  $N$-number mass shells. Shell interiors are also curved
  spacetimes matching smoothly to the exteriors through (three dimensional)
   infinitely thin shells. We found the mass ( also the charge)
   densities on each shell. The solutions obtained this way are
   free of singularities. For the case of Kastor-Traschen
   spacetimes choosing the sign of the Hubble constant $H$
   properly we showed that shells are moving away from each other
   with constant velocity.

   Shell models in higher dimensional theories seem to be also very
   interesting. In order that our method to be applicable to such
   theories one has to modify Einstein field equations with the
   inclusion of a dust matter as we did in this work where the
   mass is distributed on thin shells. In this case the shells
   are also higher dimensional. In the case of five dimensions for
   instance the thin shells are four dimensional spacetimes. Our
   work on this matter will be communicated elsewhere.

\vspace{1cm}

This work is partially supported by the Scientific and Technical
Research Council of Turkey and by the Turkish Academy of Sciences.

\end{document}